\documentclass[12pt,preprint]{aastex}

\def\Msol{\thinspace\hbox{$\hbox{M}_{\odot}$}}

\def\a4{\hsize 17.0cm \vsize 25.cm}
\newcommand{\der}[2]  { \frac{{\rm d}#1}{{\rm d}#2} }

\shorttitle{The pressure confined wind of M82-A1 supercluster}
\shortauthors{Silich et al.}

\begin{document}

\title{The pressure confined wind of the massive and compact superstar 
cluster M82-A1}

\author{
Sergiy Silich, Guillermo Tenorio-Tagle}
\affil{Instituto Nacional de Astrof\'\i sica Optica y
Electr\'onica, AP 51, 72000 Puebla, M\'exico; silich@inaoep.mx}
\and
\author{Casiana Mu\~noz-Tu\~n\'on}
\affil{Instituto de Astrof\'{\i}sica de Canarias, E 38200 La
Laguna, Tenerife, Spain; cmt@ll.iac.es}

\begin{abstract}
The observed parameters of the young superstar cluster M82-A1 and its associated compact HII region
are here shown to indicate a low heating efficiency or immediate loss, through radiative cooling, of a large fraction of 
the energy inserted by stellar winds and supernovae during the early evolution of the cluster.  This implies a bimodal
hydrodynamic solution which leads to a reduced mass deposition rate into the ISM, with a much reduced outflow velocity.
Furthermore, to match the observed parameters of the HII region associated to
M82-A1, the resultant star cluster wind is here shown to ought to be confined by a 
high pressure interstellar medium.  The cluster wind parameters, as well as the location of the reverse shock, its cooling length and the radius of
the standing outer HII region are derived analytically. All of these properties are then confirmed with a semi-analytical integration 
of the flow equations, which provides us also with the run of the hydrodynamic variables as a function of radius.  
The impact of the results is discussed and extended to other massive and young superstar clusters surrounded by a compact HII region. 
\end{abstract}

\keywords{Galaxies: star clusters ---  galaxies individual (M82) --- 
ISM: bubbles --- ISM: HII regions --- ISM: individual (M82-A1)}

\section{Introduction}

In many starburst, in interacting and in merging galaxies, a substantial 
fraction of star formation is concentrated in a number of compact ($R_{SC} \leq$ 10 pc), young (a 
few 10$^6$ yr) and massive ( $10^5 \leq M_{SC} \leq 10^6$) stellar clusters (superstar clusters; SSCs). 
These entities may represent the dominant mode of star formation in these 
galaxies (McCrady et al. 2003; Whitmore 2006) and if held by gravity they 
also represent the earliest stages of globular cluster evolution 
(Mu\~noz-Tu\~n\'on et al. 2004). Their 
large UV photon output and their powerful, high velocity gaseous outflows 
(the  star cluster winds) are now believed to be the major responsible agents 
for the large-scale structuring of the interstellar medium (ISM) and for the 
dispersal of heavy elements within  their  host galaxies and the 
intergalactic medium (IGM). 

M82-A1 is one of these SSCs in the galaxy M82 (Smith et al. 2006), with a 
half-light radius $R_{SC} = 3 \pm 0.5$~pc, an age, $\tau_{SC} = 6.4 \pm 
0.5$~Myr, an UV photon output $N^{tot} = (7.5 \pm 3.0) \times 
10^{50}$~s$^{-1}$, and a mass, $M_{SC} = (1.3 \pm 0.5) \times 10^6$\Msol \,
if one assumes an IMF with a  Salpeter slope and  0.1\Msol \, and 
100\Msol \, as the lower and upper mass cutoffs. Stellar evolution synthesis 
models (such as Starburst 99; see Leitherer et al. 1999) predict an average  
mechanical energy input rate for 
M82-A1 equal to $L_{SC} = 2.5 \times 10^{40}$ erg s$^{-1}$. 
Given such a large energy deposition rate, it seems surprising that M82-A1 
is, after such an evolution time,  still surrounded by a compact 
HII region, with a radius $R_{HII} = 4.5$pc, a density $n_{HII} \approx 1800$ 
cm$^{-3}$, a metallicity $\sim 1 - 2$ times solar 
and a total mass of only 5000 M$_\odot$, instead of having produced a 
large-scale superbubble. Note that a similar argument can be made for many 
of the SSCs embedded in the 150 pc nuclear region of M82 (see  de Grijs et al.
2001; Melo et al. 2005) as well as for a number of the extragalactic HII 
regions (Kewley \& Dopita, 2002; Dopita et al. 2005).
 
Smith et al. 2006 suggested that the HII region associated
to M82-A1 may result from the photoionization of a 4.5~pc shell 
blown into the ISM by the central cluster. However as  
the observed size of the HII 
region is not consistent with the standard bubble model (see Weaver et al. 
1977; Bisnovatyi-Kogan \& Silich 1995 and references therein), they assumed 
that the kinetic energy of the photoionized shell represents only a small 
fraction ($\sim 0.1$) of the mechanical energy, $L_{SC}$, released inside the 
cluster and that the high pressure of the ambient medium has reduced the 
expansion velocity to 30 km s$^{-1}$, the FWHM of the observed emission lines. 

This scenario, although intuitively correct, rises several questions. First, 
the spectro-photometric age of the cluster, $\tau_{SC} \approx 6.4$~Myr 
(Smith et al. 2006), is inconsistent with the kinematic age of the shell 
($\tau_{k}$) even if one assumes 30 km s$^{-1}$ as the expansion speed 
throughout its evolution, $\tau_k = R_{HII} / V_{exp} \approx 1.5 \times 
10^5$~yr. Second, if the matter injected by supernovae and stellar winds is 
accumulated inside the HII radius, $R_{HII} = 4.5$~pc, then the gas number 
density should be, $n _{HII} >  3 L_{SC} \tau_{SC} / 2 \pi \mu 
V^2_{\infty} R^3_{HII} \approx 2.7 \times 10^4$ cm$^{-3}$, where $\mu$ is 
the mean mass per ion, and a wind terminal speed  $V_{\infty} = 1000$ 
km s$^{-1}$ was assumed. This density is an order of magnitude larger than 
that found in the HII region. Note also that only a small fraction, 
$f_{HII} < 0.001$, of this gas could be photoionized by M82-A1 and thus in 
this interpretation, a good fraction of the mass supplied by supernovae 
explosions and stellar winds is hidden somewhere within the  4.5~pc volume.
These arguments imply that the cluster and HII region parameters  
are not consistent with the wind driven bubble model. 

In order to have a consistent model, here we look at the
detailed hydrodynamics of the gas reinserted by winds and supernovae within
the cluster volume. The original adiabatic  star cluster wind model, proposed 
by Chevalier \& Clegg (1985), described the various assumptions that have 
been used by all followers. Assumptions such as the equal spacing between 
sources within the SSC volume and the fact  that random collisions of the 
ejecta from nearby 
stellar winds and supernova explosions, were to result into a full 
thermalization of the matter reinserted by the evolving sources. Under such 
conditions, the pressure of the thermalized matter would exceed that of the 
surrounding  ISM and the injected gas would be  accelerated to leave the 
cluster with a high velocity, while establishing a stationary wind. For this 
to happen the gas has to acquire a particular velocity distribution, 
increasing almost linearly with radius from the stagnation point, the place 
where the gas velocity is equal to 0 km s$^{-1}$, which occurs at  the
cluster center, to the sound speed, $c_{SC}$, that should occur at the 
cluster surface. Out of the cluster, pressure gradients would allow the 
wind to reach its terminal speed ($v_\infty$ $\approx  2c_{SC}$).

The observed parameters of  SSCs (Ho 1997; Whitmore, 2003; Turner et al. 2003,
2004; Pasquali et al. 2004; Melo et al. 2005; Walcher et al. 2005;   
Mart\'\i{}n Hern\'andez et al. 2005;  Thompson et al. 2006; Larsen, 2006 
and references therein) 
led us however, to re-analyze the original adiabatic model of Chevalier
\& Clegg (1985) and realize that in the case of massive and compact star 
clusters, radiative cooling may crucially affect the internal structure 
and the hydrodynamics of the flow (Silich et al. 2004). In a more recent 
communication, Tenorio-Tagle et al. (2007) have shown that if the mechanical 
luminosity of a cluster exceeds a critical value (the threshold line), then 
a bimodal hydrodynamic solution for the matter reinserted by winds and SNe
is established.   In such cases, the densest central regions within the 
cluster volume undergo strong radiative cooling, what depletes 
the energy required to participate in the cluster wind. This also moves the 
stagnation radius, $R_{st}$, out of the cluster center and promotes the  
accumulation and the re-processing of the enclosed matter, into further 
generations of stars (Tenorio-Tagle et al. 2005). Meanwhile, the  matter 
injected between the stagnation radius and the cluster surface is still able 
to drive a stationary wind. This however, cools rapidly at a short distance 
from the cluster surface and becomes exposed to the UV radiation escaping
the cluster. W\"unsch  et al. (2007) has also shown  that 
the location of the threshold line that separates clusters evolving in the 
bimodal regime from those with the stagnation point at the star cluster 
center, which are  thus able to eject all of the deposited mass out of the 
cluster, can be well approximated by simple analytic expressions.
This is also the case for the  position of the stagnation radius, the radius 
that defines the amount of matter accumulated and ejected from a cluster in 
the bimodal regime. Making use of these relationships and our semi-analytic 
code (Silich et al. 2004), here we propose another interpretation of Smith's 
et al. (2006) observations of M82-A1, based on our bimodal solution.
Our model assumes that M82-A1 is embedded into a high pressure ISM,
the thermal pressure in the core of M82 ($P/k \sim 10^7$~cm$^{-3}$ K), which 
is much larger than that found in the disks of normal galaxies  
($P/k \sim 10^4$~cm$^{-3}$ K, see, for example, Slavin \& Cox, 1993). 
Our final suggestion is that M82-A1 is a good example of a cluster evolving 
in the catastrophic cooling regime and that a high pressure ISM is what 
rapidly confines its wind, leading to its associated compact HII region.

The paper is organized as follows. In section 2 we formulate our model,
discuss the input physics and some approximations used in 
section 3 to develop a set of analytic equations which allow us to
derive an approximate value of the heating efficiency required
by the parameters of the M82-A1 cluster and its associated HII region.
We confirm the analytic result by means of semi-analytic calculations
in section 4 where we also thoroughly discuss the structure
of the outflow driven by M82-A1. In section 5 we discuss
the possible impact  of M82-A1,  if immersed into different ISM environments. 
In section 6 we summarize and discuss our results.

\section{M82-A1 embedded into a high pressure ISM}

The properties of M82-A1 and in particular its mass and expected mechanical
energy input rate  and the size and mass of its associated 
HII region, lead almost unavoidably
to the conclusion that the cluster must be  embedded into a high pressure 
region which has managed to confine the cluster wind.  This implies a rapid 
hydrodynamic evolution which has caused the decay of the outer shock, 
inhibiting the formation of an outer shell of swept up matter and the built 
up of a superbubble (as otherwise described  in  Weaver et al. 1977).
Instead, a standing reverse shock sitting near the cluster surface, 
decelerates and thermalizes the outflow increasing its thermal pressure. 
The shocked gas then slows down while it cools and  is displaced away from 
the cluster, as fresh wind matter enters the shock.  
The shocked gas thus recombines at a small distance behind the reverse 
shock and becomes exposed to the UV radiation from the cluster, forming 
a stationary HII region. Note 
that upon cooling the shocked wind gas ought 
to  undergo a rapid condensation to counterbalance the lost of temperature 
(from the post-shock temperature to the assumed photoionization temperature 
$T_{HII} \sim $10$^4$ K) and be able to keep a pressure similar to that of 
the surrounding ISM ($P_{ISM}$). In this model the HII region associated to 
the central cluster results from the photoionization of three different
regions. The central zone within $R_{st}$ where the injected matter cools 
in a catastrophic manner. A second photoionized zone is generated within
the free wind region as this gas cools suddenly to $T_{HII} \sim 10^4$K
and finally, the shocked wind gas that has more recently being able to cool 
down from the high temperatures acquired after crossing the reverse shock. 
Other wind material that experienced earlier the same evolution is steadily  
pushed  out of the standing ionized outer shell by the newly incoming 
material, what allows for its  recombination, its further cooling by 
radiation and to be further condensed so that  its pressure remains equal to 
$P_{ISM}$. 

Major restrictions to the model arise from the pressure balance between 
$P_{ISM}$ and the wind ram pressure. This defines the position of the 
standing reverse shock, the size of the cooling distance that the shocked 
gas ought to travel before becoming exposed to the cluster UV radiation and
which together with the available number of photons, defines also the size 
of the outer HII region, which must  agree with the observations.
Here we show that to set $R_{HII}$ in agreement with the observations, the 
location of $R_{sh}$ and the size of its cooling distance imply that 
M82-A1 must be experiencing a bimodal hydrodynamic solution. 
In the bimodal regime, strong radiative cooling largely diminishes the cluster
thermal pressure within the densest central zones. This leads to the 
accumulation of the matter reinserted within the cooling volume, reducing
the amount of mass that a cluster returns to the ISM. This also results into 
a slower acceleration of the matter injected in the outer segments of the 
cluster ($r > R_{st}$), what leads to a reduced terminal speed of the 
resultant wind (see Tenorio-Tagle et al. 2007; W\"unsch  et al. 2007). As 
shown below, these considerations allow us to predict  which is the heating 
efficiency, or the fraction of the total energy input rate that is used to 
produce the wind of M82-A1 and which is the structure acquired by the outflow.
\begin{figure}[htbp]
\plotone{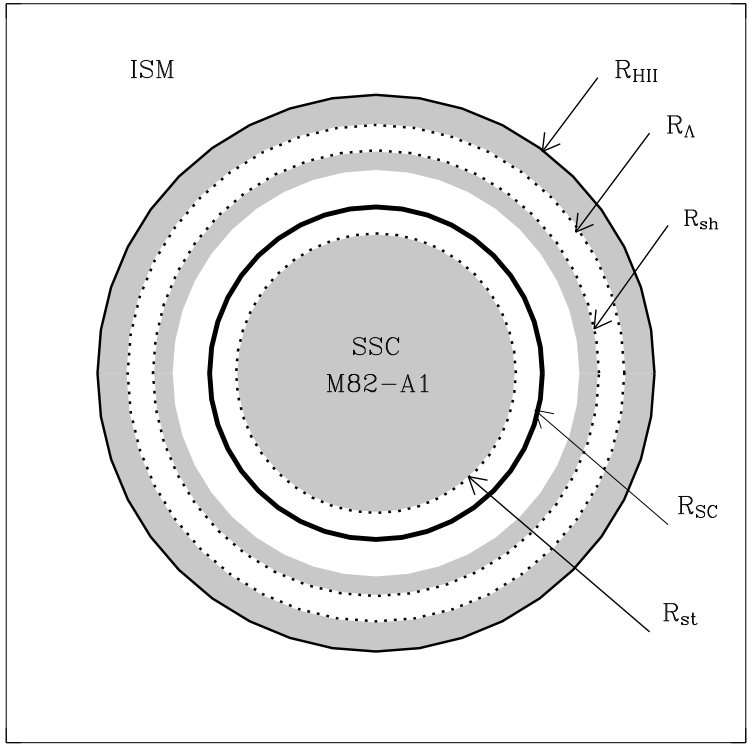}
\caption{Schematic representation of the M82-A1 HII region. The ionizing 
cluster is 
contained within the radius $R_{SC}$. $R_{st}$ is the stagnation radius 
that separates the inner cluster region, where catastrophic cooling sets in, 
from the outer zone ($R_{st} < r < R_{SC}$) where the thermalized ejecta 
remains hot and is thus able to compose a stationary wind. The star cluster 
wind however cools rapidly and becomes exposed to the UV radiation from the
cluster forming a high speed ($V_{\infty}$) HII region component. Eventually
this gas is stopped and re-heated by the reverse shock which sits at a fixed 
radius, $R_{sh}$, defined by the condition that the wind ram pressure equals 
that of the ISM. The wind processed at the reverse shock then  cools down at 
a distance $R_{\Lambda}$ to become once again target of the cluster UV 
radiation, causing in this way the outer, photoionized, thin  standing shell 
with radius $R_{HII}$ (all scales are greatly distorted). 
The model assumes an even pressure in all regions: from the ISM 
to the location of the reverse shock where the wind ram pressure equals that
of the ISM.  Regions filled with photoionized material are shaded.}
\end{figure}

To derive an analytic expression for the position of the stagnation point, 
W\"unsch et al. (2007) used the mass conservation law together 
with the fact that the density at the star cluster surface scales almost
linearly with that at the stagnation point. They found that for star clusters 
with a given radii $R_{SC}$, the location of the stagnation radius depends on 
the excess star cluster mechanical luminosity above the critical value, 
$L_{crit}$:
\begin{eqnarray}
\label{eq1a}
      & & \hspace{0.0cm}
R_{st} = R_{SC} \left[1 - 
         \left(\frac{L_{crit}}{L_{SC}}\right)^{1/2}\right]^{1/3} ,
      \\[0.2cm]
      \label{eq1b}
      & & \hspace{0.0cm}
L_{crit} = \frac{3 \pi \eta \alpha^2 \mu_i^2 R_{SC} V^4_{A\infty}}
      {2 \Lambda_{st}}
      \left(\frac{\eta V^2_{A\infty}}{2}  - \frac{c^2_{st}}{\gamma-1}\right) ,
\end{eqnarray}
where $0 < \eta < 1$ is the heating efficiency or fraction
of the deposited kinetic energy that after full thermalization is not 
immediately radiated away, and thus  is instead evenly spread within the star 
cluster volume, causing the overpressure that drives the wind (Stevens \& 
Hartwell 2003, Melioli \& de Gouveia Dal Pino  2004). Here $\alpha = 0.28$ 
is a fiducial coefficient (see W\"unsch et al. 2007), $\mu_i = 14 m_H / 11$ 
is the mean mass per ion, $V_{A\infty} = (2 L_{SC}/{\dot M}_{SC})^{1/2}$ 
is the adiabatic wind terminal speed, ${\dot M}_{SC}$ is the mass deposition
rate, $c_{st}$ and $\Lambda_{st}$ are  the speed of sound and the value of 
the cooling function at the stagnation point, respectively. Note also that 
the density and the temperature at the stagnation radius are related by the 
equation (see Silich et al. 2004; W\"unsch  et al. 2007):
\begin{equation}
      \label{eq1dd}
n_{st} = q^{1/2}_m \left[\frac{\eta V^2_{A\infty}/2 - 
           c^2_{st} / (\gamma - 1)}{\Lambda(T_{st},Z)}\right]^{1/2} , 
\end{equation}
where  $q_m = (3 {\dot M}_{SC}) / (4 \pi R^3_{SC})$ is the mass deposition 
rate per unit volume.
 
Figure 1 presents a schematic representation of the model. M82-A1 is 
contained within the inner solid line. The dashed line interior to this 
marks the location of the stagnation radius. Interior to $R_{st}$,  
cooling dominates over the heating provided by stellar winds and supernovae. 
The injected matter there cools down rapidly and is unavoidably reprocessed 
into new generations of stars supporting a low level of star formation as 
the parent cluster evolves (see Tenorio-Tagle et al. 2005). 

The outer cluster zone, $R_{st} < r < R_{SC}$, is filled with hot thermalized 
ejecta unable to radiate away all of its thermal energy within a 
dynamical time scale, and thus capable of 
moving  out of the cluster composing a stationary wind. The amount of mass 
reinserted by the  cluster wind is (W\"unsch  et al. 2007):
\begin{equation}
\label{eq1c}
{\dot M}_{out} = {\dot M}_{SC} \left(\frac{L_{crit}}{L_{SC}}\right)^{1/2} .
\end{equation}  

Outside the cluster there is a free wind region  ($R_{SC} < r < R_{sh}$) 
bound by the standing reverse shock at a radius  $R_{sh}$. The free wind
is also able to cool rapidly, loosing its thermal pressure while becoming 
easy target of the cluster UV radiation field and thus forming another
component of the HII region associated to the cluster. At the reverse 
shock the ejected material, which streams  with its terminal velocity
$V_{\infty}$, is decelerated and re-heated. The matter behind $R_{sh}$ 
would remain hot until it reaches the cooling radius $R_{\Lambda}$ 
\begin{equation}
\label{eq1d}
R_{\Lambda} = R_{sh} + L_{\Lambda} ,
\end{equation}
where the cooling length $L_{\Lambda}$ is (Franco, 1992):
\begin{equation}
\label{eq1e}
L_{\Lambda} = \frac{f_{\lambda} V_{\infty} \tau_{cool}}{4} =  
              \frac{3}{8} \frac{\mu_i}{\mu_e} 
              \frac{k f_{\lambda} T_s V_{\infty}}{n_s \Lambda(T_s)} ,
\end{equation}
where $\mu_e = 14 m_H / 23$ is the mean mass per particle in a fully
ionized plasma that contains 1 helium atom per every 10 atoms of hydrogen,
$k$ is the Boltzmann's constant, $T_s = 3 \mu_e V^2_{\infty}/16 k$ and 
$n_s = (\gamma + 1) / (\gamma - 1) \rho_w / \mu_i$ are the post-shock 
temperature and the post-shock ion density, respectively, $\Lambda(T_s)$ 
is the cooling function and $f_{\lambda}$ is a fiducial coefficient
that takes into consideration the enhance cooling promoted by the continuous 
growth of density in the post-shock region. Good agreement between formula
(\ref{eq1e}) and  numerical calculations was found for  $f_{\lambda} = 0.3$
(see below).

Outside $R_\Lambda$, the wind gas is photoionized by the UV radiation 
produced by the star cluster, forming a standing photoionized shell, with a 
pressure identical to that of the ISM.  

Note that the location of the various disturbances 
depends only on the assumed value of $P_{ISM}$ and 
on the terminal speed of the cluster wind. In this way, the density of the 
wind at the shock radius, $\rho_w$, is 
\begin{equation}
\label{eq2}
\rho_w = P_{RAM} / V^2_{\infty} ,
\end{equation}
where $P_{RAM}$, the wind ram pressure, is equal to $P_{ISM}$. In the
strongly radiative case the terminal speed of the wind,
$V_{\infty}$, falls below the adiabatic value.  Nevertheless, one can
use the adiabatic relation between $V_{\infty}$ and $c_{st}$, the sound 
speed at the stagnation point (see  Cant{\'o} et al. 2000), as a good
approximation that allows one to determine the radiative wind terminal
speed if the sound speed at the stagnation radius is known: 
\begin{equation}
\label{eq2a}
V_{\infty} = \left(\frac{2 q_{eff}}{q_m}\right)^{1/2} = 
             \left(\frac{2}{\gamma-1}\right)^{1/2} c_{st} ,
\end{equation}
where the radiative energy losses  across
the cluster were assumed to be identical to those at the stagnation point. The
effective energy deposition within  the cluster is then:
\begin{equation}
\label{eq2b}
q_{eff} = \eta q_e - n^2_{st} \Lambda_{st} = 
          \frac{2}{\gamma-1} \frac{c^2_{st}}{V^2_{A\infty}} \, q_e ,
\end{equation}
where $q_e$ is the average energy deposition rate per unit volume within  
the cluster.

\section{The analytic model}

The structure of the outflow can be derived analytically from  
a set of equations that consider: The available number of ionizing photons, 
photoionization balance, pressure confinement and the divergency of the 
wind stream:
\begin{eqnarray}
\label{eq3a}
      & & \hspace{0.0cm}
N^{out} = f_t N^{SC} ,
      \\[0.2cm]
      \label{eq3b}
      & & \hspace{0.0cm}
R_{\Lambda} = R_{HII}
\left[1 - \frac{3 N^{out}}{4 \pi \beta n^2_{HII} R^3_{HII}}\right]^{1/3}
      \\[0.2cm]
      \label{eq3c}
      & & \hspace{0.0cm}
\rho_w(R_{sh}) V^2_{\infty} = P_{ISM} ,
      \\[0.2cm]
      \label{eq3d}
      & & \hspace{0.0cm}
\rho_w(R_{sh}) = \frac{{\dot M}_{out}}{4 \pi R^2_{sh} V_{\infty}}
\end{eqnarray}
where $\beta = 2.59 \times 10^{-13}$ cm$^{-3}$ s$^{-1}$ is the recombination 
coefficient to all but the ground level, $N^{SC}$ is the total number of UV 
photons produced by the cluster and $N^{out}$ is the total number of photons 
escaping the cluster and the free wind region, and thus available to impact on 
the shocked wind gas once this has cool by radiation and has acquired by 
condensation a density $n_{HII}$. The stagnation radius, 
$R_{st}$, and  the amount of matter that a cluster returns to the ISM, 
${\dot M}_{out}$, are defined by equations (\ref{eq1a}), (\ref{eq1b}) and
(\ref{eq1c}). $f_t$ is the fraction of the star cluster ionizing radiation 
that reaches the cool shocked gas behind the standing reverse shock.
Combining equations (\ref{eq3b}), and (\ref{eq1d}), one can derive the 
expression for the reverse shock radius, $R_{sh}$:
\begin{equation}
\label{eq4a}
R_{sh} = R_{HII} \left[1 - 
         \frac{3 N^{out}}{4 \pi \beta n^2_{HII} R^3_{HII}}\right]^{1/3}
         - \frac{3}{8}\frac{\mu_i}{\mu_e} 
           \frac{k f_{\lambda} T_s V_{\infty}}{n_s \Lambda(T_s)} .
\end{equation}  
On the other hand, the position of the reverse shock is defined by the
mass conservation (\ref{eq3d}) and pressure balance 
(\ref{eq3c}) equations  and by the amount of mass that the cluster returns 
to the ISM (equation \ref{eq1c}):
\begin{equation}
\label{eq4b}
R_{sh} = \left(\frac{{\dot M}_{out} V_{\infty}}{4 \pi P_{ISM}}\right)^{1/2} =
         \frac{(4 L_{crit} L_{SC} V^2_{\infty})^{1/4}}
               {(4 \pi P_{ISM} V^2_{A\infty})^{1/2}} ,
\end{equation}  
where $V_{\infty}$ and $V_{A\infty}$ are the radiative and adiabatic
wind terminal speeds, respectively  (see W\"unsch  et al. 2007). 

Substituting equation (\ref{eq4b}) into equation (\ref{eq4a}) we obtain
a nonlinear algebraic equation which defines the heating efficiency,
$\eta$, required to match the observed parameters of the HII region 
($R_{HII}$, $N^{obs}$) associated to M82-A1, if one accounts for 
the star cluster parameters ($R_{SC}$, $L_{SC}$) and the inferred pressure 
of the ISM.
\begin{eqnarray}
      & & \hspace{0.0cm}
1 - \frac{(4 \pi P_{ISM} V^2_{A\infty} R^2_{HII})^{1/2}}
    {(4 L_{crit} L_{SC} V^2_{\infty})^{1/4}} \times
\nonumber 
\\[0.2cm]
      \label{eq5}
      & & \hspace{0.0cm}
\left[\left(1 - \frac{3 f_t N^{SC}}
{4 \pi \beta n^2_{HII} R^3_{HII}}\right)^{1/3}
      - \frac{9}{512} \frac{f_{\lambda} \mu^2_i V^5_{\infty}}
        {P_{ISM} R_{HII} \Lambda(T_s)}\right] = 0 .
\end{eqnarray}  

If the adiabatic wind terminal speed, $V_{A\infty}$, is known, the only
free parameter in equation (\ref{eq5}) is $0 < f_t < 1$, the fraction of 
UV photons able to reach the outer HII region. One can then solve 
equation (\ref{eq5}) by iterations for a given value of $f_t$. 

The solution of equation (\ref{eq5}) shows that the heating efficiency
which is required in order to fit the observed structure of the M82-A1 HII 
region is small, $\eta \approx 4.65$\%, and does not depend significantly
on the adopted value of $f_t$. Figure 2 displays the structure of the outflow produced by M82-A1 as 
well as the dimensions of its associated outer HII region, calculated under 
the assumption that $V_{A\infty} = 1000$ km s$^{-1}$. 
The low value of the heating efficiency results into a large stagnation
radius ($R_{st} \approx 2.94$pc, the lower solid line in Figure 2) 
which implies also that a large fraction 
of the matter supplied by stellar winds and supernovae is to remain bound
to the cluster and thus only the small amount deposited between $R_{st}$ and 
$R_{SC}$ would conform the cluster wind while expanding approximately  
with 200 km s$^{-1}$. The outflow is strongly decelerated and re-heated
at the reverse shock (4.24pc $\le R_{sh} \le 4.35$pc, the second solid line 
from the bottom in Figure 2) and then cools down rapidly to
form an outer shell of photoionized matter depleting the photons that escape
the cluster ($N^{out}$). Note that the thickness of this outer shell is very
small, and becomes even smaller  as one assumes a smaller  number of photons escaping the SSC and the cluster wind. 
\begin{figure}[htbp]
\plotone{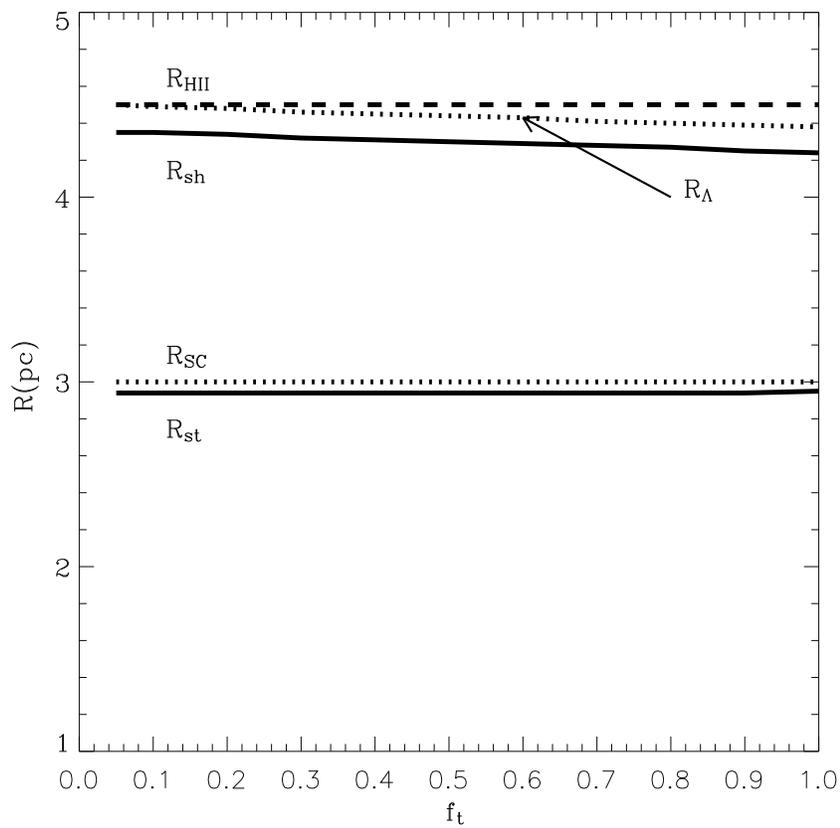}
\caption{The analytic model of M82-A1 and its associated  HII region. 
The various lines mark the location of different hydrodynamical disturbances 
as a function of $f_{t}$. From bottom to top these represent the location 
of the stagnation radius ($R_{st}$), the cluster radius ($R_{SC}$),
the reverse shock location ($R_{sh}$), the cooling radius, $R_{\Lambda}$ 
and the outer radius of the standing HII region ($R_{HII}$).}
\end{figure}

\section{The semi-analytic model}

\begin{figure}[htbp]
\vspace{17.5cm}
\includegraphics{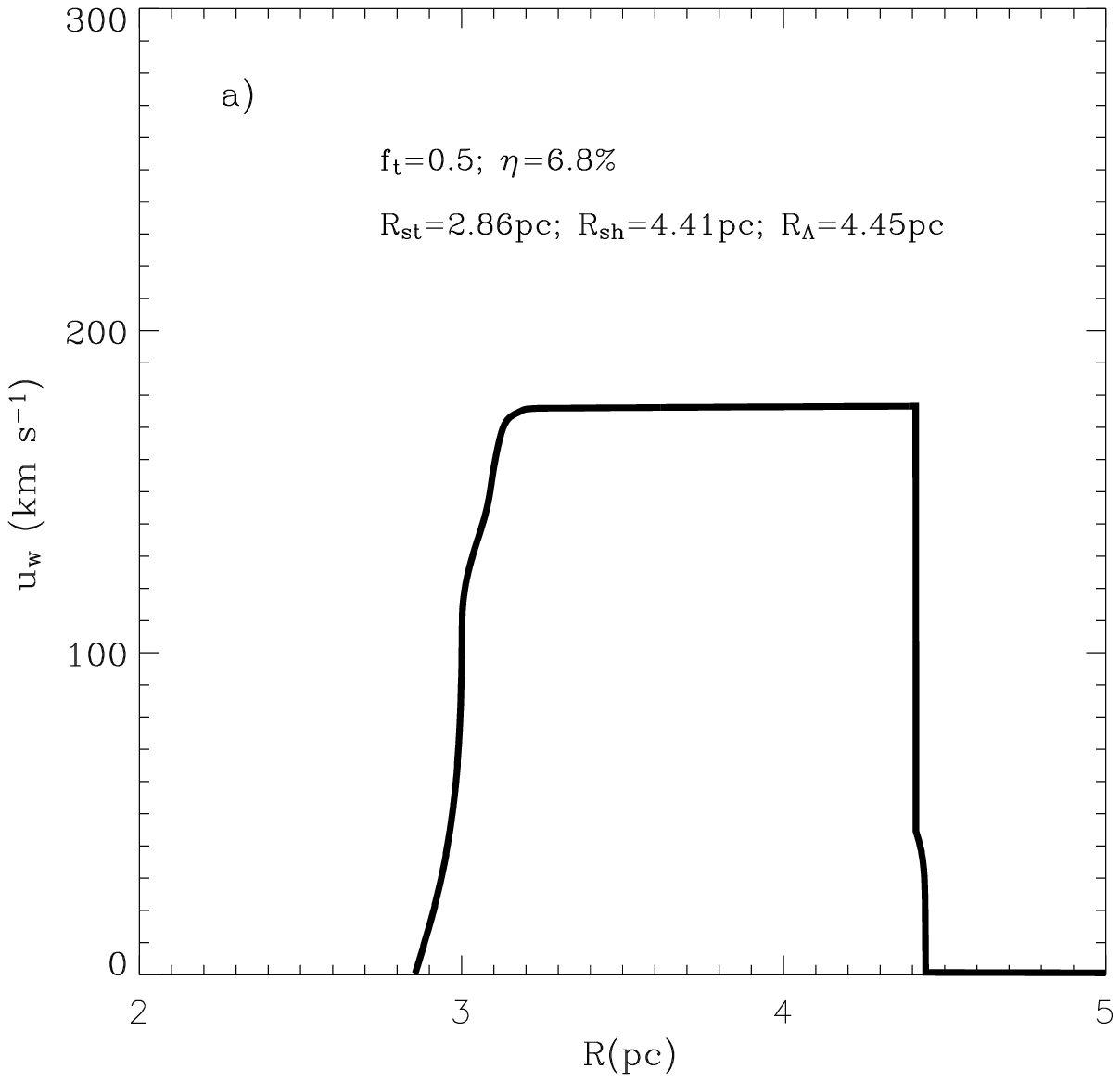}
\includegraphics{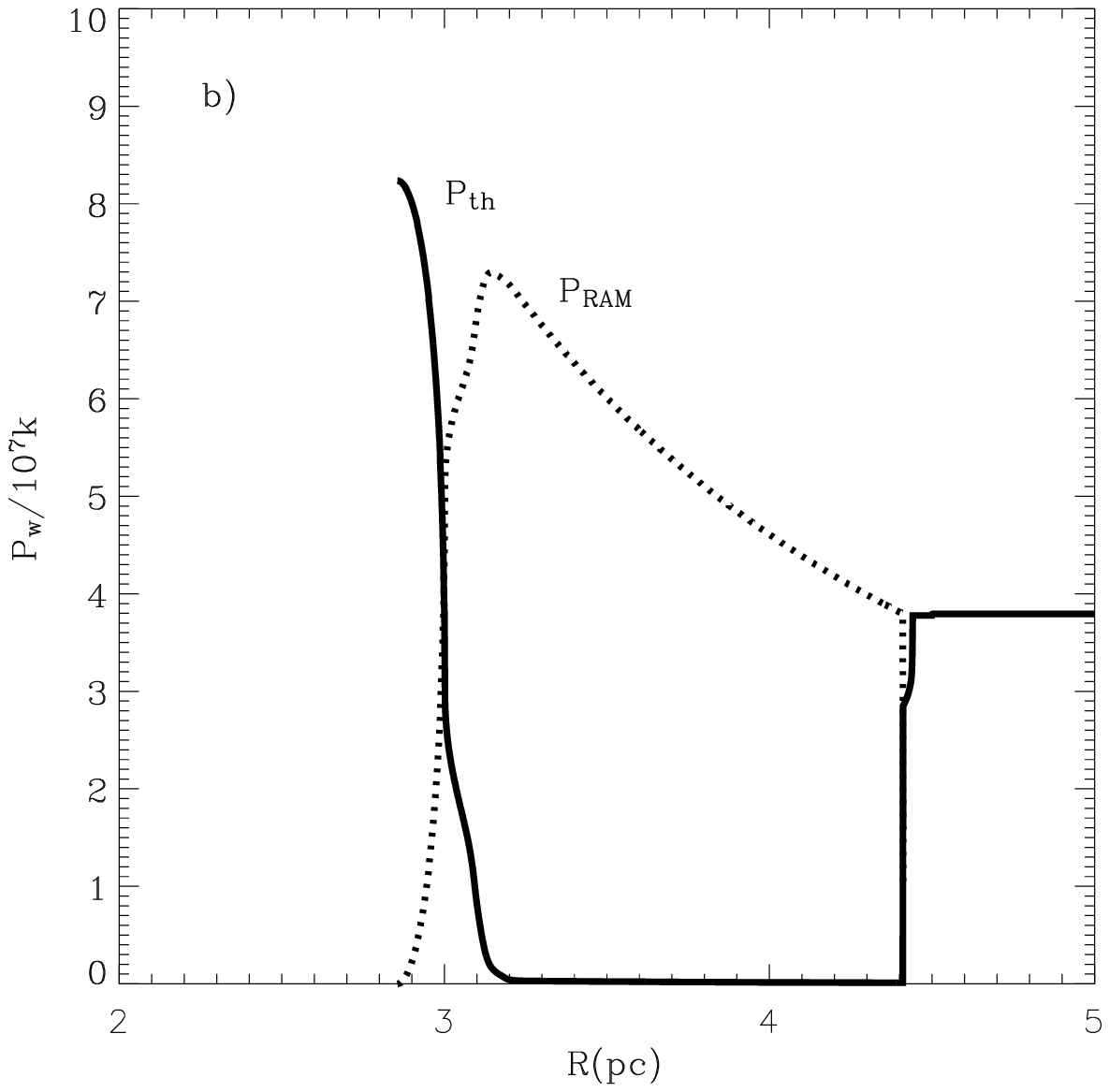}
\includegraphics{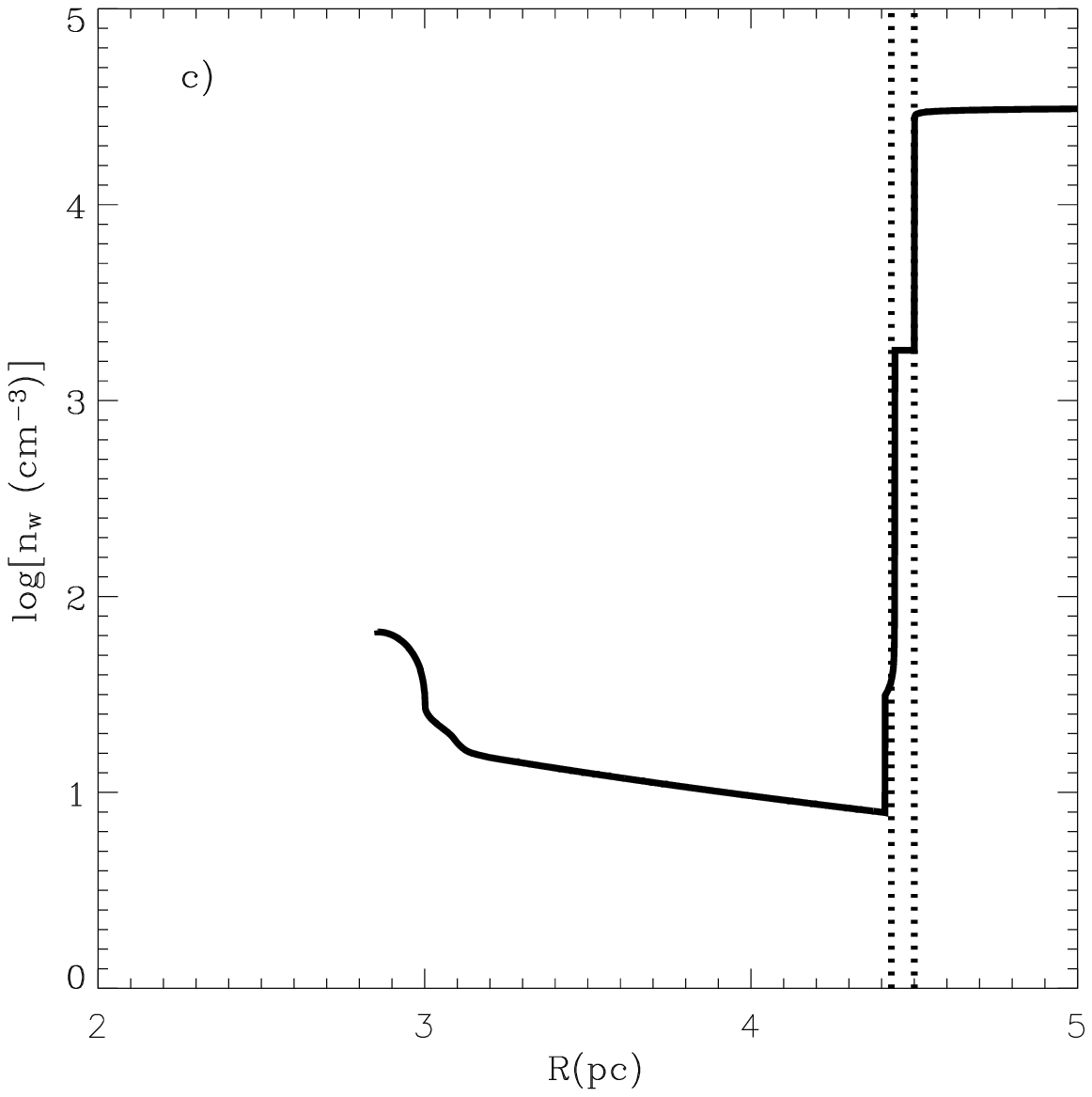}
\includegraphics{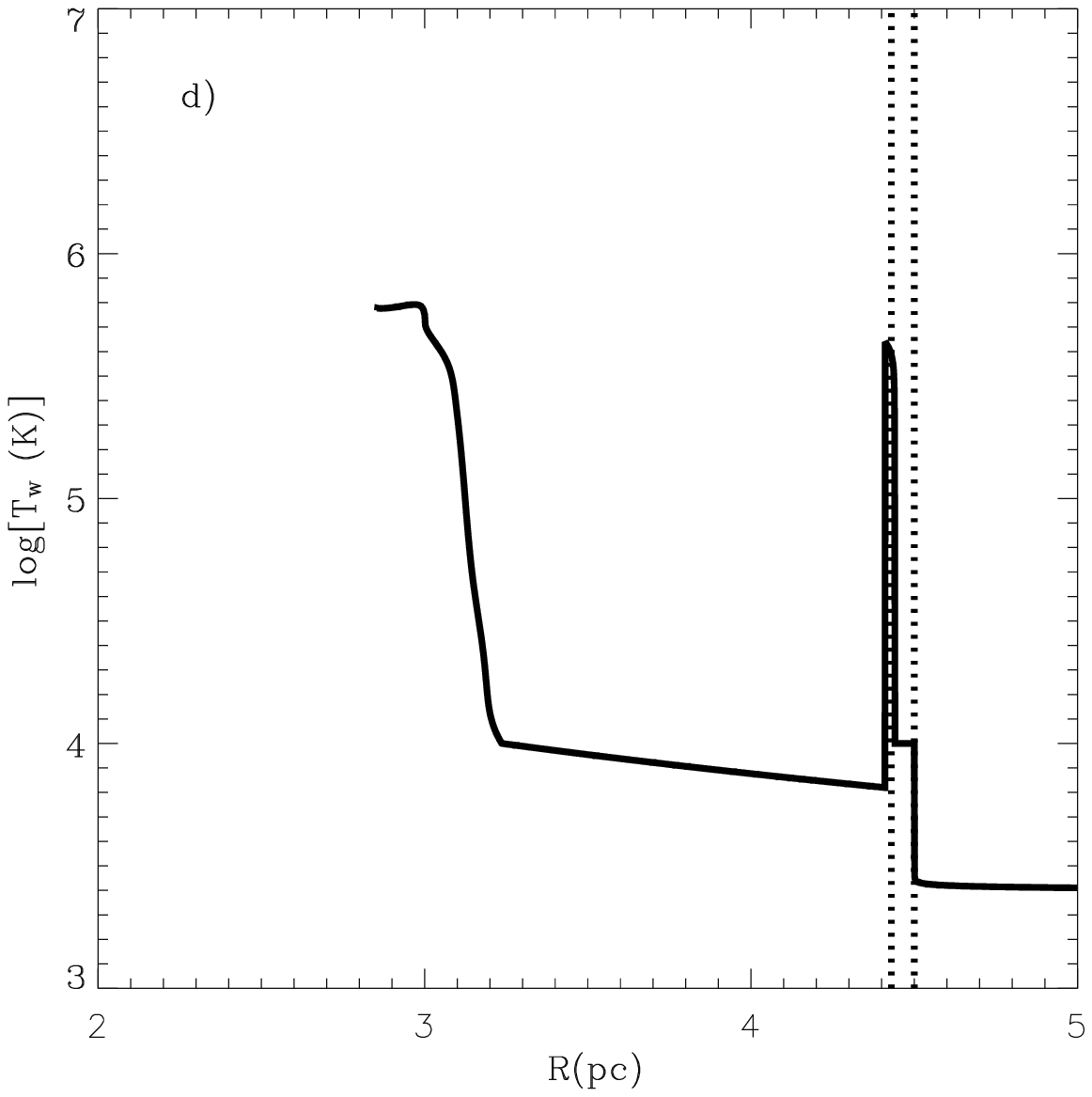}
\caption{The semi-analytic model of M82-A1 and its associated  HII region.
Panels a -  d present the distributions of the velocity,
thermal (solid line) and ram (dotted line) pressures, density and 
temperature, respectively. The size of the standing outer HII region
is indicated in panels c and d by the two vertical dotted lines. The 
calculations assumed an $f_t = 0.5$, $Z = Z_{\odot}$ and an 
adiabatic wind terminal speed, $V_{A\infty} = 1000$ km s$^{-1}$.}  
\end{figure}

To corroborate our results, here we use our semi-analytic code (see Silich 
et al. 2004) that accurately calculates the divergent spherically-symmetric
outflow beyond $R_{st}$, accounting for strong radiative cooling. 

We use first the trial heating efficiency $\eta$ to calculate the position 
of the stagnation point as it was 
suggested in Tenorio-Tagle et al. (2007) and then integrate the
equations of mass, momentum and energy conservation outwards,
knowing that at some distance from the cluster the standing
reverse shock sets in. We stop the integration at the reverse 
shock and use then the Rankine-Hugoniot conditions
to calculate the thermal pressure and the velocity of the plasma 
behind the shock front:
\begin{eqnarray}
      \label{eq6a}
      & & \hspace{0.0cm}
P_2 = P_w \left(\frac{2 \gamma M^2_1 - (\gamma - 1)}{\gamma + 1}\right) ,
\\[0.2cm]
      \label{eq6b}
      & & \hspace{0.0cm}
u_2 = u_w \left(\frac{\gamma-1}{\gamma+1} + \frac{2}{\gamma+1}
      \frac{1}{M^2_1}\right) ,
\end{eqnarray}  
where $P_w$ and $u_w$ and $P_2$ and $u_2$ are the thermal pressure and the 
velocity of the outflow ahead and behind the reverse shock, respectively. 
$M_1 = u_w/c_w$ is the Mach number ahead of the shock front, and
$\gamma = 5/3$ is the ratio of specific heats. $P_2$ and $u_2$ together 
with the mass conservation law, ${\dot M}_{out} = 4 \pi \rho_w u_w R^2_{sh}$, 
define the initial conditions behind the reverse shock for the set of 
main equations outside of the cluster (see Silich et al. 2004):
\begin{eqnarray}
      \label{eq7a}
      & &         \hspace{-3.0cm}
\der{u_w}{r}  = \frac{1}{\rho_w} \frac{(\gamma-1) r Q + 
              2 \gamma u_w P_w}{r (u_w^2 - c_s^2)} ,
      \\[0.2cm]   \label{eq7b}
      & & \hspace{-3.0cm}
\der{P_w}{r} = - \frac{{\dot M}_{sc}}{4 \pi r^2} \der{u_w}{r} ,
      \\[0.2cm]   \label{eq7c}
      & & \hspace{-3.0cm}
\rho_w = \frac{{\dot M}_{sc}}{4 \pi u_w r^2} .
\end{eqnarray}
The run of the hydrodynamical variables in the outer part of the flow was 
obtained by integrating equations (\ref{eq7a})-(\ref{eq7c}) from the 
reverse shock radius outwards.

The thermal pressure, $P_w$, and the wind expansion velocity, $u_w$, ahead
of the shock front depend on the star cluster parameters and on the radius
of the reverse shock, $R_{sh}$. Therefore the set of initial conditions 
(\ref{eq6a}) - (\ref{eq6b}) contains two model parameters: the value of 
the heating efficiency, $\eta$, and the position of the reverse shock, 
$R_{sh}$. We iterate $\eta$ and $R_{sh}$ until the conditions
\begin{eqnarray}
      \label{eq8a}
      & & \hspace{0.0cm}
R_{\Lambda} = R_{HII} 
\left[1 - \frac{3 N^{out}}{4 \pi \beta n^2_{HII} R^3_{HII}}\right]^{1/3} ,
      \\[0.2cm]
      \label{eq8b}
      & & \hspace{0.0cm}
P_{HII} = P_{ISM} ,
\end{eqnarray}
are fulfilled. Here $R_{HII} = 4.5$pc is the observed radius of the 
M82-A1 HII region, $P_{HII} = k n_{HII} T_{HII}$ and $T_{HII} = 10^4$K are 
the thermal pressure and the temperature of the ionized gas in the standing 
outer shell, respectively.

Figure 3 presents the results of the calculation assuming $f_t = 0.5$, e.g. 
when only half of the ionizing photons reach the outer standing shell.
The expansion velocity of the outflow grows from zero km s$^{-1}$ at
the stagnation radius to rapidly reach its terminal value, 
$V_{\infty} \approx 180$ km s$^{-1}$, and then drops when the free wind
reaches the reverse shock at $r \approx 4.4$pc (panel a in Figure 3). 
The gas decelerates rapidly at the reverse shock and then as it cools down 
it is further condensed. Thus in the case of the pressure confined 
wind,  the reverse shock separates gas that flows supersonically
in the free wind region from gas moving away with a subsonic 
velocity.      
The thermal pressure (see panel b) drops initially when the gas is 
accelerated to reach its terminal speed, and remains at its lowest value as 
a consequence of radiative cooling. $P_{th}$ is restored after crossing the 
reverse shock, where the ram pressure in the free wind reaches a balance
with the thermal pressure in the surrounding interstellar medium,
and finally after photoionization of the dense outer layer. The density 
decreases as $r^{-2}$ in the free wind region. It is then compressed 
at the shock, reaches, upon condensation induced by radiative cooling, 
the HII region value and finally the maximum value in the neutral layer 
outside of the outer photoionized shell (Panel c in Figure 3). 
Figure 3 also shows that in the bimodal regime the 
temperature of the outflowing matter drops rapidly outside of the cluster. The
free wind is then re-heated at the reverse shock and cools down again
to form the outer shell whose inner skin is photoionized by the Lyman
continuum escaping from the M82-A1 cluster. Note that in our calculations
the temperature of the free wind (between the star cluster
surface and the reverse shock) drops to  $10^4$K and thus a broad
($u_w \approx 180$km s$^{-1}$), low intensity emission line component is 
expected  from this region.

Certainly, the analytic model is unable to show the detailed structure of 
the outflow. Nevertheless, the analytic heating efficiency, stagnation 
radius, reverse shock and cooling radii are in reasonable agreement 
with the results obtained  from the semi-analytic model: $\eta = 6.8\%$, 
$R_{st} = 2.86$pc, $R_{sh} = 4.41$pc and $R_{\Lambda} = 4.45$pc, if
the fraction of ionizing photons reaching the outer HII region is $f_t = 0.5$.

\section{M82-A1 in different environments}

The theory developed in the previous sections is based on the assumption
that the HII region detected around the M82-A1 is a standing, photoionized 
shell of shocked wind matter, confined by the high pressure of the ISM.
A major implication of the results above is that clusters with properties 
(size, mass and age) similar to those of M82-A1 would have a low heating 
efficiency. Here we assume such a cluster to be embedded into different 
ISM environments and workout the properties of the resultant pressure 
confined remnant. 
 
The different pressure in the ambient ISM does not  affect the 
distribution of matter within the cluster. It would not affect the position 
of the stagnation radius, nor the density, temperature and velocity of the 
ejected plasma. A different ISM pressure would modify only the outer 
structure of the outflow shifting the position of the reverse shock. This is  
because of the pressure equilibrium condition: $P_{RAM} = 
\rho_w(R_{sh}) V^2_{\infty} = P_{HII} = P_{ISM}$. Using equations (\ref{eq1c})
and (\ref{eq3d}) one can then obtain the position of the reverse shock
for M82-A1 sitting in different interstellar environments:
\begin{equation}
\label{eq6}
R_{sh} = \left(\frac{{\dot M}_{out} V_{\infty}}{4 \pi P_{ISM}}\right)^{1/2} =
         \left(\frac{{\dot M}_{SC} V_{\infty}}{4 \pi P_{ISM}}\right)^{1/2}
         \left(\frac{L_{crit}}{L_{SC}}\right)^{1/4} .
\end{equation}  
One can then find the outer radius of the HII region from equations
(\ref{eq1d}), (\ref{eq1e}) and (\ref{eq3b}), knowing that the density 
in the ionized outer shell is a linear function of the pressure in the 
ambient interstellar medium:
\begin{equation}
\label{eq7}
n_{HII} = P_{ISM} / k T_{HII} ,
\end{equation}  
where the temperature of the photoionized shell, $T_{HII}$, is set through 
photoionization and therefore it does not depend on $P_{ISM}$. Equation 
(\ref{eq6}) indicates also that there exists the critical 
interstellar pressure, $P_{crit}$: 
\begin{equation}
\label{eq8}
P_{crit} = \frac{{\dot M}_{SC} V_{\infty}}{4 \pi R^2_{SC}}
           \left(\frac{L_{crit}}{L_{SC}}\right)^{1/2} 
\end{equation}  
If the pressure in the ambient ISM exceeds this  critical value, 
$P_{ISM} > P_{crit}$, the cluster would not have the sufficient power to 
drive a cluster wind. In this case all matter deposited by stellar winds 
and SNe would remains buried within the cluster.
\begin{figure}[htbp]
\plotone{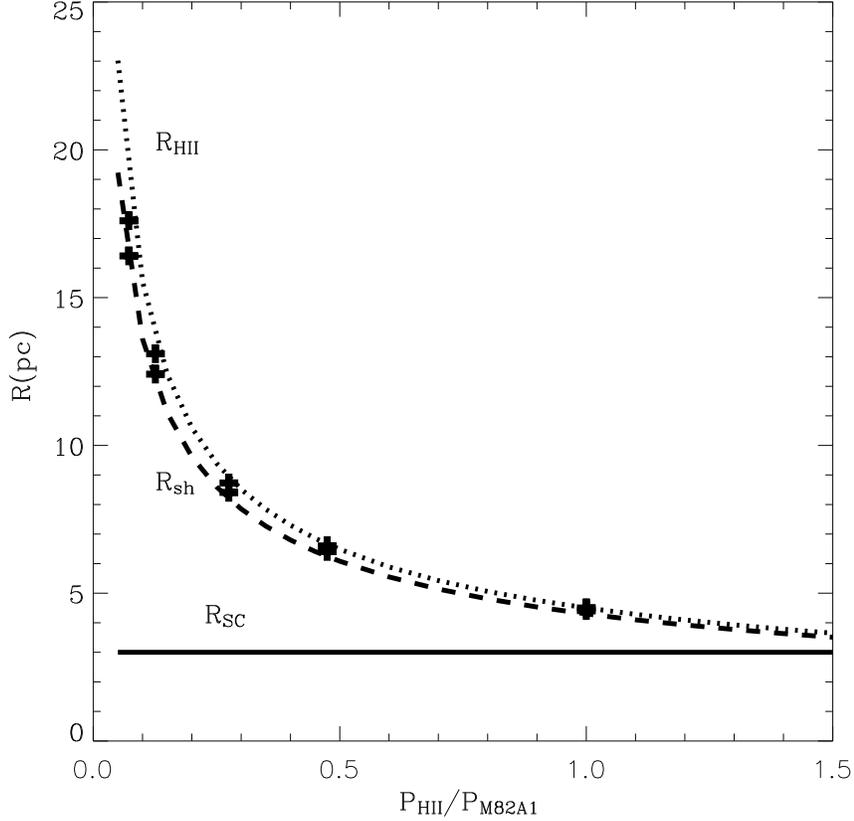}
\caption{The structure of pressure confined winds into different ISM 
environments. We have assumed the star cluster parameters derived by 
Smith et al. (2006) for M82-A1 and a heating efficiency, $\eta = 4.6\%$, as 
derived in section 3. The solid line marks the star cluster radius which 
is equal to the half-light radius of M82-A1. The dashed and dotted lines
mark the position of the reverse shock, $R_{sh}$, and of the outer HII region, 
$R_{HII}$, respectively, as a function of  
$P_{ISM}$ normalized to that found by Smith et al. (2006) in the M82-A1 
HII region. The cross symbols represent the results of the semi-analytic
calculations and also mark the positions of the shock front and 
the outer edge of the HII region (lower and upper symbols, respectively). 
The fraction of the star cluster ionizing radiation 
able to photoinize the outer shell is $f_t = 0.5$ in both, the analytic 
and semi-analytic calculations.}
\end{figure}

Figure 4 shows the structure of the pressure confined remnants that 
M82-A1 would produce in different interstellar environments. 
Our results show that the outer ionized shell is very thin.
Also that the radius of the ionized shell shrinks towards the star cluster  
if one considers a pressure in the interstellar medium that approaches the 
critical value (equation \ref{eq8}) and grows as the considered pressure in 
the interstellar medium drops. 
The analytic results (dashed and dotted lines in Figure 4) show reasonable 
agreement with those obtained by means of semi-analytic calculations (cross
symbols in Figure 4).

\section{Results and Discussion}

The results from our  analytic formulation to a pressure confined wind in 
agreement with those obtained with our semi-analytical code, imply that 
M82-A1 is a massive and compact cluster with a low heating efficiency. This 
implies a completely different result from what one would expect from an 
adiabatic model, as a low heating efficiency leads to a bimodal hydrodynamic 
solution and with it to a low mass deposition rate into the ISM  with a much 
reduced outflow velocity.

Furthermore, to match the observed parameters of M82-A1 and its associated 
HII region, our results lead also to a high pressure environment able to 
confine the cluster wind by setting a reverse shock close to the star cluster 
surface. In this way the outflow is thermalized, what leads to temperatures 
near the top of the interstellar cooling curve and thus to a rapid cooling of 
the strongly decelerated outflow. The wind becomes then target of the cluster 
UV radiation, composing a narrow standing outer shell of photoionized gas 
with the shocked wind matter that continuously traverses the reverse shock.

Our calculations lead to three HII region components associated to M82-A1:  
A central component where the deposited matter cools catastrophically and 
does not participate in the wind. A section of the free wind region which, 
upon expansion,  is also able to cool rapidly. And the outer stationary shell 
of shocked wind matter, which defines the observed size of the associated HII 
region. Note that, given the densities in the latter component, this ought to 
be the one producing the most intense emission lines, while    
the second component is to produce a broad (~ 2 $V_{\infty}$) 
low intensity emission line component, similar to that detected in 
NGC 4214-1 (see Ho \& Filippenko 1996), able to enhance, as in the case of 
M82-A1, the width of the observed lines.
  
Our two methods of solution have also shown good agreement when considering 
the wind of M82-A1 being pressure confined by  an ISM with different pressures.

Note that in the case of  very  young SSCs (a few Myrs old) their mass is 
usually derived from the intensity of their optical emission lines, emanating 
from their associated compact HII regions and thus through a measure of the 
available UV photon flux, an assumed IMF and stellar synthesis models 
(Melo et al. 2005, Smith et al. 2006).
In the light of the models here presented, this implies both that young SSCs 
are embedded into a high pressure ISM and also that their heating efficiency 
ought to be rather small for their winds to be pressure confined in the 
immediate neighborhood of the clusters, and thus present an associated 
low mass and compact HII region.

Note also that the two different populations of massive and compact clusters 
(Gyr old and very young clusters) present a similar range of masses as well 
as a similar size distribution. This can be explain 
if most of the mass reinserted by stellar winds and supernova explosions
during the early evolutionary stages (which may amount to 30 per cent of the 
$M_{SC}$ released during the first 50 Myr of evolution) is not returned to 
the ISM but rather re-processed in situ into further stellar generations. 
This would allow  SSCs to keep most of their mass (and size) as they age and 
are able to avoid their dispersal. This also implies that a large 
fraction of the low mass stars presently observed in old compact and massive
stellar clusters might have formed from the matter injected by massive
stars. In this respect our bimodal model provides the hydrodynamical ground 
to a self-enrichment scenarios. In these, for example, the abundance 
anomalies observed in globular clusters may result from the enrichment of 
the proto-stellar gas during the earlier stages of their evolution
(see, for example, Prantzos \& Charbonnell, 2006; Decressing et al. 2007
for a comprehensive discussion of different aspects of this problem).
Our solution also suggests that the metallicity of the compact HII regions
associated with the intermediate age (5~Myr - 10~Myr) massive clusters should
be super-solar and variable with time.

\acknowledgments 
We thank our anonymous referee for valuable comments and suggestions.
This study has been supported by CONACYT - M\'exico, research grant 
47534-F and AYA2004-08260-CO3-O1 from the Spanish Consejo Superior de
Investigaciones Cient\'\i{}ficas.




\end{document}